\documentclass[aps,twocolumn,showpacs,preprintnumbers,amsmath,amssymb, prb, reprint]{revtex4-1}
\usepackage{amsmath}
\usepackage{stmaryrd}
\usepackage{txfonts}
\usepackage{amssymb}
\usepackage{mathrsfs}
\usepackage{graphicx}
\usepackage{dcolumn}
\usepackage{epsfig}
\usepackage{hyperref}
\usepackage{graphics}
\usepackage{color}
\usepackage{wrapfig}
\usepackage{enumerate}
\usepackage{amssymb}
\usepackage{bm}

\newcommand{\sgn}{\text{sgn}}

\makeatother

\begin{document}

\title{Spin-orbit coupling and  electronic charge effects in Mott insulators}

\author{Shan Zhu}
\affiliation {Department of Physics, Zhejiang Institute of Modern Physics, Zhejiang
University, Hangzhou 310027, People's Republic of China}
\author{You-Quan Li}
\email{yqli@zju.edu.cn}
\affiliation {Department of Physics, Zhejiang Institute of Modern Physics, Zhejiang
University, Hangzhou 310027, People's Republic of China}

\author{Cristian D. Batista}
\email{cdb@lanl.gov}
\affiliation {Theoretical Division, T4 and CNLS, Los Alamos National Laboratory, Los Alamos, New Mexico 87545, USA}

\begin{abstract}
We derive the effective charge- and current-density operators for the strong-coupling limit of a single-band Mott insulator in the presence of spin-orbit coupling and show that the spin-orbit contribution to the effective charge density leads to novel  mechanisms for multiferroic behavior. In some sense, these mechanisms are the electronic counterpart of the ionic-based mechanisms, which have been proposed for explaining the electric polarization induced by spiral spin orderings. The new electronic mechanisms are illustrated by considering cycloidal and proper screw magnetic orderings on sawtooth and kagome lattices. As for the isotropic case, geometric frustration is crucial for achieving this purely electronic coupling between spin and charge degrees of freedom. 
\end{abstract}
\maketitle

\section{introduction}

The Mott insulator is a paradigmatic example of a state of matter which cannot be understood without including the crucial role of the electron-electron Coulomb interaction. Valence electrons  of a half-filled band localize in their atoms because of the large Coulomb energy cost $U$ of double occupying the same atomic orbital. A low-energy spin degree of freedom arises from the fact that each atomic orbital is mainly occupied by a {\it single} electron.   Spin degrees of freedom interact with each other via virtual processes through
double-occupied states which arise from a partial electronic delocalization: Electrons have finite hopping amplitude $t$ between neighboring atomic orbitals. This simple mechanism leads to  a large spectrum of exotic phases that range from usual magnetic orderings, such as the Neel antiferromagnetic phase, to rather exotic spin liquid states.~\cite{Zapf14,Balents10} 

Understanding the magnetic properties of different Mott insulators has been one of the main goals of condensed matter physics over the past decades.  However, the recent interest in multiferroic compounds and magnetoelectric (ME) effects generated large efforts for understanding the charge effects, which are still present in Mott insulators. This interest in the interplay between magnetism and ferroelectricity in magnetic ferroelectrics, or multiferroics, was triggered by the giant ME effects observed in some geometrically frustrated 
magnets~\cite{Kimura2003,Hur2004,Kimura2005,Lawes2005,Taniguchi2006}. In contrast to  conventional ME  effects, in which the induced  electric polarization is  linear or bilinear in the applied magnetic field ${\bf H}$, the giant ME effects of certain frustrated magnets are not a smooth function of ${\bf H}$ because the magnetic  field induces an electric phase transition.

Most efforts for understanding the microscopic mechanisms behind giant ME effects  focused on the ionic displacements  induced  by certain magnetic orderings (magnetostriction). However, as was later recognized, purely electronic charge effects also exist  in Mott insulators.~\cite{PhysRevB.78.024402} Indeed, the possibility of having a spontaneous  electronic polarization in strongly correlated insulators had been already pointed out in the context of the ionic Hubbard model~\cite{Batista04,Aligia05}.
Interestingly enough, the particle-hole symmetry of Mott insulators implies that  purely electronic contributions to the electric polarization only exist in frustrated systems, i.e., systems in which it  is possible to close a loop with an {\it odd} number of hopping amplitudes.~\cite{PhysRevB.78.024402,Lin13}
Purely electronic charge effects were originally studied for SU(2) invariant models, i.e., models which do not include  spin-orbit coupling (SOC).~\cite{PhysRevB.78.024402} In this isotropic limit, the electronic charge redistribution is associated with a modulation of  bond operators of the form $\langle {\bf S}_{j}  \cdot  {\bf S}_{l} \rangle$~\cite{Kamiya12}. This modulation can exist already in collinear spin structures, implying that non-collinear spin ordering is  not a prerequisite for inducing electronic polarization. 

The perturbative effect of a weak SOC on the effective change density operator of strongly coupled Mott insulators was derived recently  to model the  optical conductivity of these materials.~\cite{arxiv.1113} Here we focus on the influence of the SOC on the electronically induced ME effects.  Moreover, because the spin-orbit interaction is not weak for heavy magnetic ions, such as $4d$ and $5d$ transition metals, or lanthanide ($4f$) and actinide ($5f$) elements, it is important to analyze the charge effects induced by SOC beyond the weak-coupling regime. 

The ionic-based mechanisms rely either on the magnetostriction induced by certain bond orderings (both the bond angle and the bond length are modulated by a periodic change in the scalar product of two neighboring spin moments $\langle {\bf S}_{j}  \cdot  {\bf S}_{l} \rangle $) or on the spin-orbit interaction, which triggers an ionic displacement for spiral spin orderings. Spiral spin states are characterized by a nonzero vector chirality $\langle {\bf S}_{j}  \times  {\bf S}_{l}\rangle \neq 0$ between neighboring spins. The so-called "inverse Dzyaloshinskii-Moriya (DM)" mechanism produces a net electric dipole proportional to  ${\bf e}_{jl} \times \langle {\bf S}_{j}  \times  {\bf S}_{l}\rangle$ (${\bf e}_{jl}$ is the relative unit vector between spins $j$ and $l$)~\cite{Katsura05,PhysRevB.73.094434,PhysRevLett.96.067601}. The polarization is induced by the displacement $\delta {\bf x}$ of a third ion (with charge $q_I$) away from the bond center ( this ion mediates the superexchange interaction between spins $i$ and $j$) .  The induced DM interaction, ${\bf D}_{jl} \propto \delta {\bf x} \times {\bf e}_{jl}$, lowers the magnetic energy by an amount  ${\bf D}_{jl} \cdot \langle {\bf S}_{j}  \times  {\bf S}_{l}\rangle$, which is linear in $\delta {\bf x}$. Because the elastic energy cost is quadratic in $\delta {\bf x}$, the electric polarization $q_I \delta {\bf x}$ is finite as long as $\langle {\bf S}_{j}  \times  {\bf S}_{l}\rangle \neq 0$. 

The inverse DM mechanism is only active for cycloidal spiral orderings: ${\bf Q} \perp \langle {\bf S}_{j}  \times  {\bf S}_{l}\rangle$, where  ${\bf Q}$ is the wave vector  of the spiral. 
Arima recently proposed an alternative metal-ligand hybridization mechanism  which allows for  a net electric polarization induced by proper screw spiral spin ordering (${\bf Q} \parallel \langle {\bf S}_{j}  \times  {\bf S}_{l}\rangle$) as long as the lattice symmetry is low enough~\cite{JPSJ.76.073702,Jia2007}. This mechanism has been observed in different compounds {\cite{PhysRevB.77.052401,PhysRevLett.103.237601,PhysRevLett.106.167206,PhysRevB.86.060403,Seki13042012,PhysRevB.87.014429,PhysRevB.87.100402}.

A question then arises as to whether it is possible to have spiral orderings inducing a purely electronic contribution to the  net electric polarization. Here we demonstrate that this is indeed possible, as long as the Hamiltonian is frustrated and it includes a finite SOC. 
To achieve this goal we consider the strong-coupling (large $U/t$) limit of Mott insulators and derive effective low-energy operators for observables associated with the charge degrees of freedom. 
In Sec.~\ref{hfh} we introduce the half-filled Hubbard model with finite SOC and derive general expressions for the effective charge- and current-density operators. 
In Sec.~\ref{wsocl}, we consider the weak SOC limit of the effective charge-density operator to explain how the finite SOC leads to a new 
mechanism for electronic ferroelectricity induced by spiral ordering. We consider the cases of cycloidal and proper-screw spiral orderings on frustrated 
kagome and sawtooth lattices. Finally, we compare these novel electronic mechanisms against the well-known inverse DM and Arima's mechanisms for spiral ordering and summarize the results in Sec.~\ref{conc}.

\section{Half-filled Hubbard model with spin orbit coupling \label{hfh}}

\subsection{Effective Hamiltonian in the strong coupling limit}

We consider one-band Hubbard model at half filling in the presence
of SOC~\cite{Moriya,PhysRevLett.69.836}: 
\begin{equation}
H=-\sum_{\left\langle j l \right\rangle} [\mathbf{c}_{j}^{\dagger}({\tau}_{jl}+\mathbf{d}_{jl}\cdot {\bm \sigma}) \mathbf{c}_l
+\mathrm{H.c.} ]+U\sum_{j}  n_{j \uparrow}n_{j \downarrow},
\label{eq:Hamiltonian}
\end{equation}
with
\[ \mathbf{c}_{j}^{\;}= \left( \begin{array}{c}
c_{j \uparrow}^{\;}  \\
c_{j \downarrow}^{\;}  \end{array} \right),
\;\;\;\;
\mathbf{c}_{j}^{\dagger}= \left( c_{j \uparrow}^{\dagger} ,
c_{j \downarrow}^{\dagger}   \right).
\]
Here $c_{j \alpha}^{\dagger}$ ($c_{j \alpha}$) denotes the creation
(annihilation) of an electron at site $j$ with spin $\alpha=\{ \uparrow,\downarrow \}$,
$n_{j\alpha}=c_{j\alpha}^{\dagger} c_{j\alpha}$ is the electron
number operator with spin $\alpha$ at site $j$, and $t_{jl}$ and
$\mathbf{d}_{jl}$ are the coefficients of the hopping matrix  between sites $j$ and
$l$ in the basis defined by the identity and the three Pauli matrices $\sigma^{\nu}\,(\nu=x,y,z)$. 
The scalar product $\mathbf{d}_{jl}\cdot {\bm \sigma}$ represents the  SOC. 
The second term describes the  Coulomb repulsion between electrons on the
same orbital. 

For convenience, we introduce $\mathcal{A}_{jl}\equiv t_{jl} e^{i\theta_{jl}\mathbf{n}_{jl}\cdot {\bm \sigma}}$,
where $-\pi < \theta_{jl} \leq \pi$, $t_{jl} \cos\theta_{jl}=\tau_{jl}$ , $t_{jl} \sin\theta_{jl}=\left|\mathbf{d}_{jl}\right|$
and $\mathbf{d}_{jl}=i\mathbf{n}_{jl}\left|\mathbf{d}_{jl}\right|$ ($\mathbf{d}_{jl}$ is pure imaginary and $|\mathbf{n}_{jl}|=1$). 
Therefore, 
\begin{equation}
\tan{\theta_{ij}} = \frac{\left|\mathbf{d}_{ij}\right|}{\tau_{ij}}, \;\; t_{ij} = \frac{\tau_{ij}}{\cos{\theta_{ij}}},
\end{equation}
and $\sgn{(\theta_{jl})}=\sgn{(t_{jl})}$. We note that $\theta_{jl}=\theta_{lj}$, $t_{jl}=t_{lj}$ and $\mathbf{n}_{jl} = - \mathbf{n}_{lj}$.
The effect of a $U(1)$ vector potential $\mathbf{A}(\mathbf{r})$
is to add a global phase to the hopping matrix: $t_{jl}=-texp[-i\frac{e}{\hbar}\int^{{\bf r}_l}_{{\bf r}_j} \mathbf{A}(\mathbf{r})\cdot d\mathbf{l}]$,
where $t$ is real and the integral is along the hopping path (Peierls
substitution). In contrast, the SOC leads to an SU(2) rotation of the hopping matrix in each bond  (Wilson line): $\mathcal{A}_{jl}= P\{exp[-i \int^{{\bf r}_l}_{{\bf r}_j} \mathbf{\sigma}^{\nu}(\mathbf{A}^{\nu}\cdot d\mathbf{l})]\}$,
where $P$ is the path-ordering operator and we are using Einstein's convention of summation over  repeated index $\nu$. 
%We have assumed that the nearest neighbor hopping path $\left\langle ij\right\rangle $ is infinitesimal. 
We note that $\mathcal{A}_{jl}^{\dagger}=\mathcal{A}_{lj}$ because 
the Hamiltonian is Hermitian. The  hopping term can then be rewritten in the more compact form
\begin{equation}
H_{t}=-\sum_{\left\langle jl \right\rangle } t_{jl} \mathbf{c}_{j}^{\dagger} \mathcal{A}{}_{jl}  \mathbf{c}_{l}+\mbox{H}.\mbox{c}.\label{eq:hopping}
\end{equation}
In the following we consider the strong coupling limit ($U\gg t$) and expand in the small parameter $t/U$.
The $t=0$  ground space is spanned by the $2^{N}$ states with one electron per site, where $N$ is the total number of sites. The
massive spin degeneracy is lifted to order $t^2/U$. This result can be  obtained by applying  an adequate canonical transformation,
\begin{equation}
H'\equiv e^{\mathcal{S}}He^{-\mathcal{S}}\equiv H+\left[\mathcal{S},H\right]+\frac{1}{2!}\left[\mathcal{S},\left[\mathcal{S},H\right]\right]+\cdots.
\end{equation}
The linear in $t$ terms mix the $t=0$ ground space with states including one double
occupied site. These terms can be eliminated by  choosing
a generator $\mathcal{S}$ that satisfies: $H_{t}+[\mathcal{S},\, H_{U}]=0$,
where $H_{U}$ is the Hubbard or interacting term of~\eqref{eq:Hamiltonian}. The hopping terms can be divided into three contributions~\cite{PhysRevB.37.9753,PhysRevB.41.2565},
\begin{equation}
H_{t}=T_{1}+T_{0}+T_{-1},\label{eq:hoopingT}
\end{equation}
where
\begin{align*}
T_{0} & =-\sum_{\left\langle j l \right\rangle ,\alpha\alpha'} t_{jl} \{n_{j \bar{\alpha}}c_{ j \alpha}^{\dagger}\left(\mathcal{A}_{jl}\right)_{\alpha\alpha'}c_{l \alpha'}n_{l \bar{\alpha'}} +h_{j \bar{\alpha}} c_{j\alpha}^{\dagger}\left(\mathcal{A}_{jl}\right)_{\alpha\alpha'} c_{l\alpha'}h_{l\bar{\alpha'}}\},\\
T_{1} & =-\sum_{\left\langle jl \right\rangle ,\alpha\alpha'}  t_{jl} n_{j \bar{\alpha}}c_{j\alpha}^{\dagger}\left(\mathcal{A}_{jl}\right)_{\alpha\alpha'} c_{l \alpha'}h_{l \bar{\alpha'}},\\
T_{-1} & =-\sum_{\left\langle jl \right\rangle ,\alpha\alpha'} t_{jl} h_{j\bar{\alpha}}c_{j\alpha}^{\dagger}\left(\mathcal{A}_{jl}\right)_{\alpha\alpha'}c_{l\alpha'}n_{l\bar{\alpha'}},
\end{align*}
$h_{j \alpha}=1-n_{j \alpha}$ and $\bar{\alpha}$ denotes the spin orientation opposite to $\alpha$.
The hopping  terms $T_{1}$
and $T_{-1}$   change the number of double-occupied
sites, whereas $T_{0}$ conserves this number.
To eliminate the hopping terms among states with different numbers $n_d$
of double-occupied sites up to order $t^{k}$ ($E_{0}\propto n_{d}U$), we  introduce $\mathcal{S}=i\mathcal{S}^{(1)}+i\mathcal{S}^{(2)}+\cdots+i\mathcal{S}^{(k)}$
where $i\mathcal{S}^{(k)}$ is proportional to $t^{k}$.  One can verify  that $[T_{q},H_{U}]=-qUT_{q}$ with $q=(1,-1,0)$. Through
a recursive scheme we obtain 
\begin{eqnarray}
i\mathcal{S}^{(1)} & = & U^{-1}(T_{1}-T_{-1}),\nonumber \\
i\mathcal{S}{}^{(2)} & = & U^{-2}([T_{1},T_{0}]+[T_{-1},T_{0}]),\nonumber \\
i\mathcal{S}^{(3)} & = & U^{-3}\{[[T_{1},T_{0}],T_{0}]-[[T_{-1},T_{0}],T_{0}]\nonumber \\
 &  & +\frac{1}{4}[[T_{1},T_{0}],T_{1}]-\frac{1}{4}[[T_{-1},T_{0}],T_{-1}]\nonumber \\
 &  & +\frac{2}{3}[T_{1},[T_{1},T_{-1}]]-\frac{2}{3}[T_{-1},[T_{-1},T_{1}]]\},
\end{eqnarray}
to third order in $t/U$.  

Any low energy effective operator is obtained from 
\begin{equation}
{\tilde O} =P_{s} e^{\mathcal{S}}Oe^{-\mathcal{S}} P_{s},
\label{canon}
\end{equation}
where $P_{s}$ is the projector into the subspace $S_0$ of  states with no double-occupied sites (ground space for  $t=0$). At half filling, any effective
operator can be expressed as a function of the spin operators $\mathbf{S}_{j}=\frac{1}{2} {\bf c}_{j}^{\dagger} \mathbf{\sigma} {\bf c}_j$
because each state of $S_0$ is fully characterized by its spin configuration. 
For instance, up to an irrelevant constant term, the effective low-energy spin Hamiltonian to second order in $t$ is a tilted Heisenberg model~\cite{Moriya,PhysRevLett.69.836,PhysRevB.84.205123},
\begin{equation}
{\tilde H}=\sum_{\left\langle jl \right\rangle } \frac{4t^{2}_{jl}}{U} \mathbf{S}^{\dagger}_{j}  \mathcal{J}_{jl} \mathbf{S}_{l}.\label{eq:effHami}
\end{equation}
Here we have introduced the notation
\[ \mathbf{S}_{j}= \left( \begin{array}{c}
S^x_j  \\
S^y_j \\
S^z_j \end{array} \right),
\;\;\;\;
\mathbf{S}_{j} ^{\dagger}= \left( S^x_j ,S^y_j, S^z_j  \right).
\]
$\mathcal{J}_{jl}$ is the exchange tensor on bond $\langle j l \rangle$ that is a 
function of $\theta_{jl}$ and $\mathbf{n}_{jl}$,
\begin{equation}
 \mathcal{J}_{jl} \mathbf{S}_{l}  =\cos\left(2\theta_{jl}\right)\mathbf{S}_{l}+\sin\left(2\theta_{jl}\right)(\mathbf{S}_{l}\times\mathbf{n}_{jl})+2\sin^{2}\left(\theta_{jl}\right)\mathbf{n}_{jl}(\mathbf{n}_{jl}\cdot\mathbf{S}_{l}).
 \label{eq:DMtrand}
\end{equation}
By replacing this expression in~\eqref{eq:effHami}, we obtain
\begin{equation}
{\tilde H} = {\tilde H}_{\rm iso} + {\tilde H}_{\rm asm} +{\tilde H}_{\rm sm},
\end{equation}
with
\begin{eqnarray}
{\tilde H}_{\rm iso} &=& \sum_{\left\langle jl\right\rangle } \frac{4t^{2}_{jl}}{U}
\cos2\theta_{jl} \; \mathbf{S}_{j}\cdot\mathbf{S}_{l}
\nonumber \\
{\tilde H}_{\rm asm} &=& \sum_{\left\langle jl\right\rangle } \frac{4t^{2}_{jl}}{U} 
\sin2\theta_{jl} \; \mathbf{n}_{jl}\cdot(\mathbf{S}_{j}\times\mathbf{S}_{l})
\nonumber \\
{\tilde H}_{\rm sm} &=& \sum_{\left\langle jl\right\rangle } \frac{8 t^{2}_{jl}}{U} 
\sin^{2}\theta_{jl} \; (\mathbf{n}_{jl}\cdot\mathbf{S}_{j})(\mathbf{n}_{jl}\cdot\mathbf{S}_{l})
 \label{eq:effHami2}
\end{eqnarray}
$H_{\rm iso}$ is the usual isotropic Heisenberg  interaction with exchange constants $J_{jl}= 4t^{2}_{jl}/U$. The SOC generates $H_{\rm asm}$ and $H_{\rm sm}$,   which correspond to the
exchange anisotropy. $H_{\rm asm}$ is the anti-symmetric exchange anisotropy known as DM interaction.
${\tilde H}_{\rm sm}$ is the symmetric exchange anisotropy.

\subsection{Effective charge- and current-density operators}

The charge-density operator,
\begin{equation}
n_{j} = \mathbf{c}_{j}^{\dagger}  \mathbf{c}_{j},
\label{dens}
\end{equation}
and the current density operator,
\begin{equation}
{\bf I}_{jl}=\frac{i {\bf e}_{jl} t_{jl}}{\hbar } ( \mathbf{c}_{l}^{\dagger} \mathcal{A}{}_{lj}  \mathbf{c}_{j} -\mathbf{c}_{j}^{\dagger} \mathcal{A}{}_{jl}  \mathbf{c}_{l}),
\label{cdens}
\end{equation}
satisfy a continuity equation ${\partial_t n_j} = (i/\hbar) [H,n_j] = - \nabla \cdot {\bf I} ({\bf r}_j)$, where $ \nabla \cdot {\bf I} ({\bf r}_j)$ is the divergence of the current density operator on the lattice at the  site $j$ and ${\bf e}_{jl} = ({\bf r}_{l}  -  {\bf r}_{j})/|{\bf r}_{l}  -  {\bf r}_{j}| $. This continuity equation simply reflects the fact that $H$ conserves the total charge: $[H, \sum_j n_j] =0$. We will derive now  the effective charge- and current-density operators in the presence of SOC. This derivation will reveal the
interplay between the charge and the spin degrees of freedom in Mott insulators. As was explained in Ref.~\onlinecite{PhysRevB.78.024402}, the lowest nontrivial 
contribution to the effective charge- and current-density operators arises from loops of three sites (smallest possible loop). 
Therefore, this contribution is of third order in the hopping amplitude.
The SOC allows to flip the spin in the hopping process and opens the possibility of having charge redistribution 
even on triangles with three parallel spins. Such a charge redistribution is absent without SOC because of the
Pauli exclusion principle. 

\begin{figure}
\includegraphics[scale=0.4]{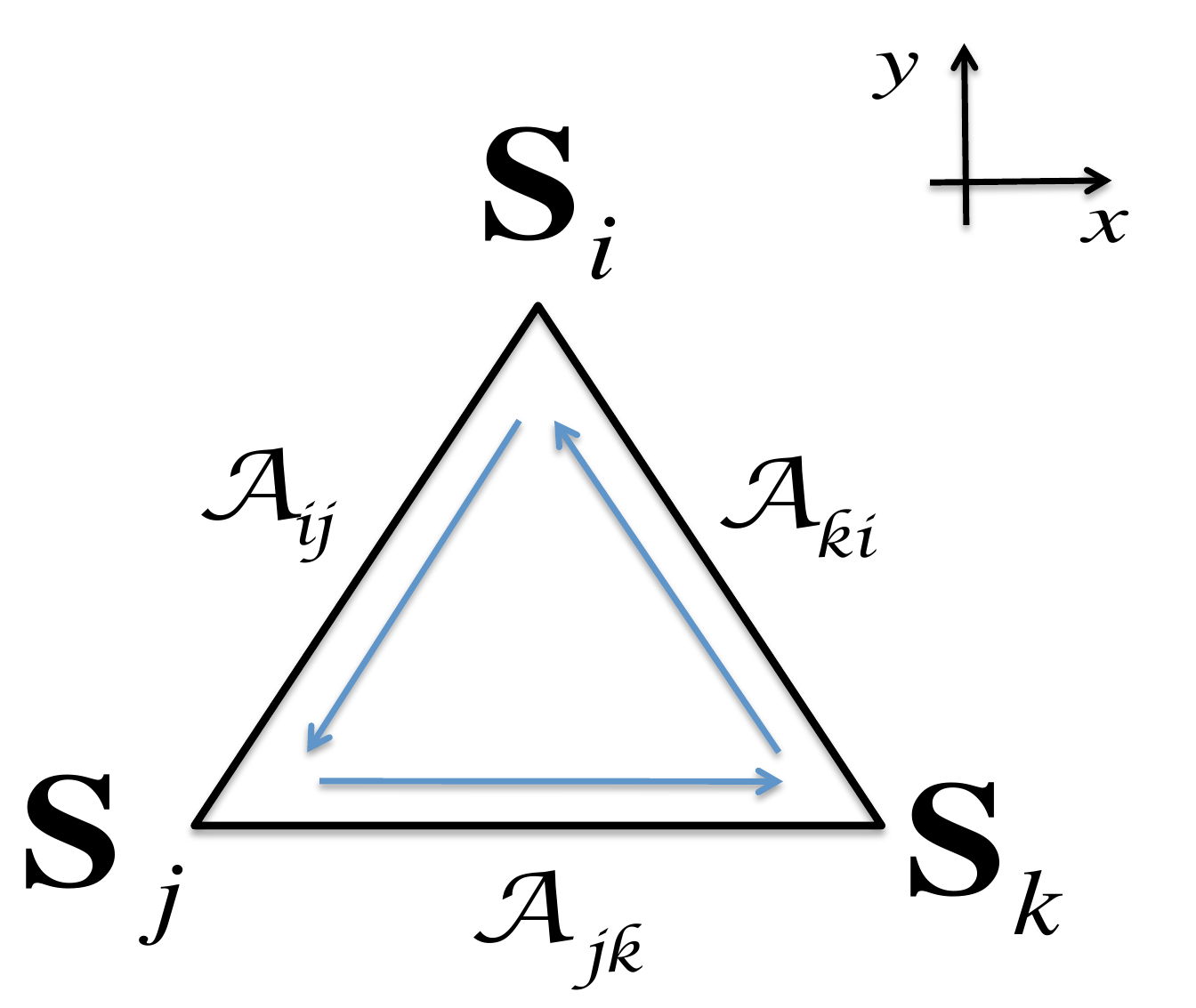}
\caption{\label{fig:SOCsche} Schematic of the hopping terms in a closed triangular loop in the presence of SOC. The SOC makes the matrix $\mathcal{A}$ different from a multiple of the identity in each bond. We use a counterclockwise convention to define the Wilson loop or SU(2) flux on the plaquette: $e^{-i\theta_{ijk}\mathbf{n}_{ijk}\cdot\mathbf{\sigma}}\equiv\mathcal{A}_{ij}\mathcal{A}_{jk}\mathcal{A}_{ki}$. In the absence of net $\mathrm{SU}(2)$ flux, we can
 simultaneously transform  the three  $\mathcal{A}$ matrices into matrices that are proportional to the identity by changing the local reference frame (applying 
 local rotations) in the spin space. This is not possible when there is a net flux, i.e., $e^{i\theta_{ijk}\mathbf{n}_{ijk}\cdot\mathbf{\sigma}}$ differs from the identity.}
\end{figure}

The deviation from one of the effective charge-density operators ($\delta {\tilde n}_{i}= {\tilde n}_{i}-1$) on the triangle   $(ijk)$  shown in Fig. \ref{fig:SOCsche} is
\begin{equation}
\delta {\tilde n}_{i}= 8 \frac{t_{ij} t_{jl} t_{li}}{U^{3}}\sum_{j,k} (\cos\theta_{ijk} X_{i,jk}-\sin\theta_{ijk}\mathbf{n}_{ijk}\cdot Y_{i,jk})
\end{equation}
 where the summation is over all possible pairs $j,k$ that close a triangular loop $ijk$ with site $i$.
 Note that for a triangle $1-3$,  there are two different contributions to the sum for $i=1$: ($j=2$,$k=3$), and ($j=3$,$k=2$).
 The operators that appear in the sum are as follows:
\begin{eqnarray}
 X_{i,jk} &=& (\mathbf{S}_{i}-\mathcal{J}_{ij}  \mathbf{S}_{j})\cdot( \mathcal{J}_{ik} \mathbf{S}_{k})
\nonumber \\
{\bf Y}_{i,jk} &=& (\mathbf{S}_{i}-\mathcal{J}_{ij} \mathbf{S}_{j})\times(\mathcal{J}_{ik} \mathbf{S}_{k}). 
\end{eqnarray} 
The phase $\theta_{ijk}$ and the unit vector $\mathbf{n}_{ijk}$ are  defined by  $e^{-i\theta_{ijk}\mathbf{n}_{ijk}\cdot\mathbf{\sigma}}\equiv\mathcal{A}_{ij}\mathcal{A}_{jk}\mathcal{A}_{ki} =exp[-i\ointclockwise_{ijk} \mathbf{A}^{[m]} {\bm \sigma}^{[m]}d\mathbf{l}]$
for the counterclockwise convention shown in Fig. \ref{fig:SOCsche}. 
$\theta_{ijk}$ denotes the magnitude of $\mathrm{SU}(2)$ flux in a triangle $ijk$, and $\mathbf{n}_{ijk}$ denotes the direction of the flux in
spin space. The
 Wilson loop is equal to the identity if the counterclockwise integral is equal to the clockwise integral. To obtain the electric polarization we need
to express the charge density of the three sites of the triangle in the same "reference frame" ($\theta_{ijk}$, $\mathbf{n}_{ijk}$). In
the following we will consider the charge redistribution of the single triangle shown in Fig.~\ref{fig:SOCsche} and we  will choose $e^{-i\theta_{ijk} \mathbf{n}_{ijk}\cdot\mathbf{\sigma}}\equiv\mathcal{A}_{ij}\mathcal{A}_{jk}\mathcal{A}_{ki}$. The charge density on each site is given by: 
\begin{eqnarray}
\delta {\tilde n}_{i} &=& 8 \frac{t_{ij} t_{jk} t_{ki}}{U^{3}} [ \cos{\theta_{ijk}} \; (X_{i,jk} + X_{i,kj}) 
\nonumber \\
 &-& \sin{\theta_{ijk}}  \; \mathbf{n}_{ijk}\cdot({\bf Y}_{i,jk} - {\bf Y}_{i,kj}) ],  
 \nonumber \\
\delta {\tilde n}_{j} &=& 8 \frac{t_{ij} t_{jk} t_{ki}}{U^{3}} [\cos{\theta_{ijk}} \; (-2X_{i,jk} 
\nonumber \\
&+& X_{i,kj})+\sin{\theta_{ijk}} \; \mathbf{n}_{ijk} \cdot(2{\bf Y}_{i,jk}+{\bf Y}_{i,kj}) ],
\nonumber \\
\delta \tilde{n}_{k} &=& 8 \frac{t_{ij} t_{jk} t_{ki}}{U^{3}} [\cos{\theta_{ijk}} \; (X_{i,jk}- 2 X_{i,kj}) 
\nonumber \\
&-& \sin{\theta_{ijk}} \; \mathbf{n}_{ijk}\cdot( {\bf Y}_{i,jk}+ 2 {\bf Y}_{i,kj}) ].\label{eq:chargedis}
\end{eqnarray}
We note that charge conservation holds in the single triangular plaquette: $\delta n_{1}+\delta n_{2}+\delta n_{3}=0$. We also note that the charge density operator is explicitly invariant under a global  rotation
 (both the spin and ${\bf n}$ are rotated simultaneously) and under a time-reversal transformation.
 
\begin{figure}
\includegraphics[width=7cm]{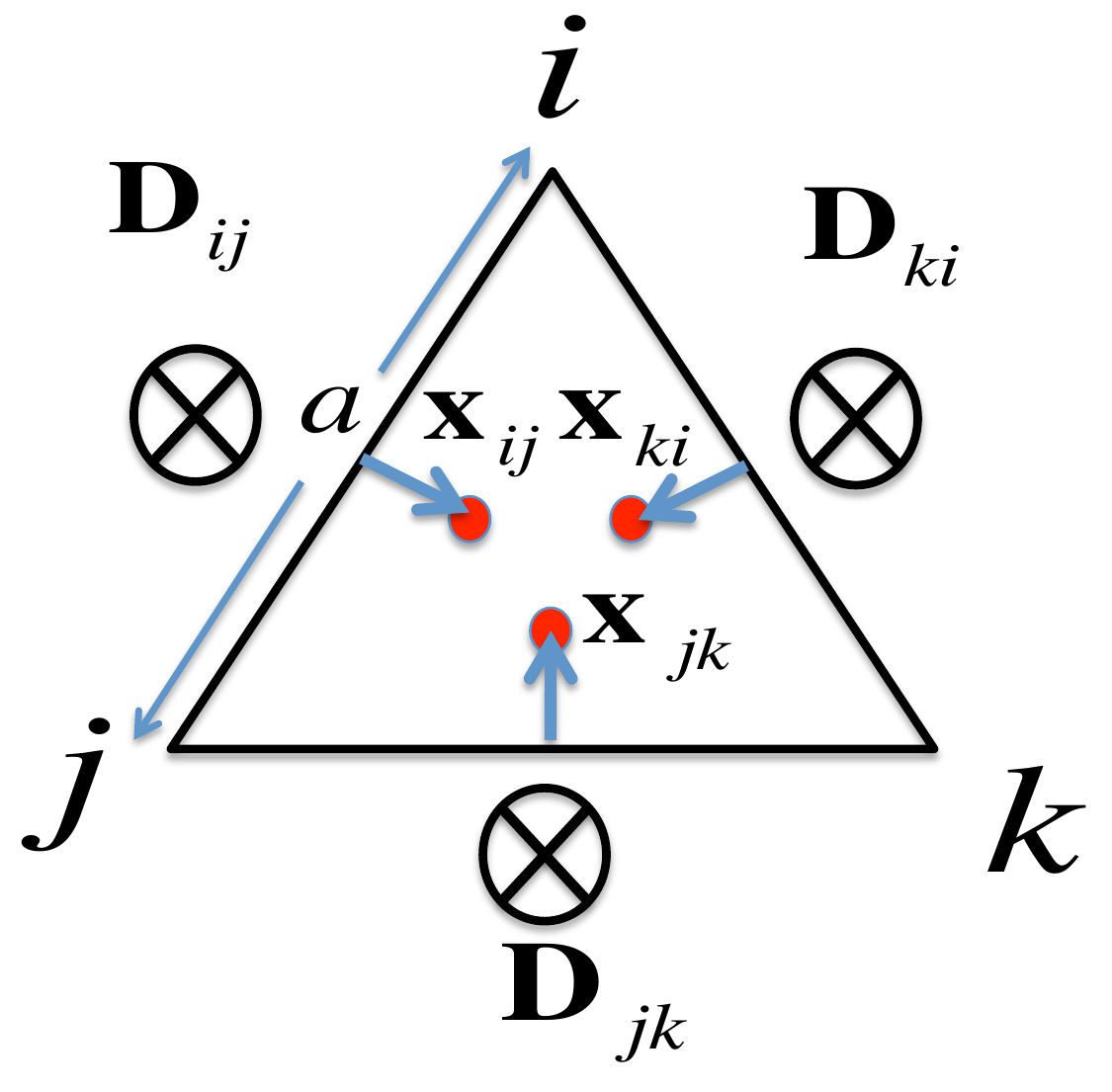}
\caption{\label{fig:oxy} Illustration of one particular realization of the DM vector with the same magnitude on each bond 
and  antiparallel to the $z$ axis. The atoms that appear inside the triangle mediate the hopping between two sites. The DM vector on 
 bond $\langle ij \rangle$ is ${\bf D}_{ij} \propto {\bf x}_{ij} \times {\bf r}_{ij} = - (2 J \theta/3) {\bf e}_z $. It is then clear that  ${\bf D}_{ij} = {\bf D}_{jk} ={\bf D}_{ki} $}
\end{figure} 

Based on the charge redistribution in Eqs.~\eqref{eq:chargedis},
the electric polarization in an equilateral triangle is given by the following expressions:
\begin{equation}
\tilde{\mathbf{P}} = \tilde{\mathbf{P}}_{ij}+\tilde{\mathbf{P}}_{jk}+\mathbf{\tilde{P}}_{ki}
\label{eq:px}
\end{equation}
with
\begin{eqnarray}
\tilde{\mathbf{P}}_{ij} &=& 8ea\frac{t_{ij}t_{jk}t_{ki}}{U^3}[\cos\theta_{ijk}\mathbf{S}_i\cdot\mathcal{J}_{ij}\mathbf{S}_{j}
\nonumber \\
&+&\sin\theta_{ijk}\mathbf{n}_{ijk}\cdot(\mathbf{S}_i\times\mathcal{J}_{ij}\mathbf{S}_j)]
 \; (\mathbf{e}_{jk}-\mathbf{e}_{ki}),
\nonumber \\
\tilde{\mathbf{P}}_{jk} &=& 8ea\frac{t_{ij}t_{jk}t_{ki}}{U^3}
[\cos\theta_{ijk}\mathcal{J}_{ij}\mathbf{S}_{j}\cdot\mathcal{J}_{ik}\mathbf{S}_{k}
\nonumber \\
&+&\sin\theta_{ijk}\mathbf{n}_{ijk}\cdot(\mathcal{J}_{ij}\mathbf{S}_j\times\mathcal{J}_{ik}\mathbf{S}_{k})] \; (\mathbf{e}_{ki}-\mathbf{e}_{ij}),
\nonumber \\
\tilde{\mathbf{P}}_{ki} &=& 8ea\frac{t_{ij}t_{jk}t_{ki}}{U^3}
[\cos\theta_{ijk}\mathcal{J}_{ik}\mathbf{S}_{k}\cdot\mathbf{S}_i
\nonumber \\
&+&\sin\theta_{ijk}\mathbf{n}_{ijk}\cdot(\mathcal{J}_{ik}\mathbf{S}_k\times\mathbf{S}_i)] (\mathbf{e}_{ij}-\mathbf{e}_{jk}),
\label{eq:py}
\end{eqnarray}
 where $a$ is the bond length (see Fig.~\ref{fig:oxy}) and $-e$ is the electron charge. 
 
The effective current-density operator on the bond $\langle ij \rangle$ is  obtained by replacing ${\cal O}$ with the charge density operator ${\bf I}_{ij}$ in  Eq.~\eqref{canon}, 
 \begin{eqnarray}
 \tilde{\mathbf{I}}_{ij} &=& 24\frac{ {\bf e}_{ij} }{\hbar}\frac{t_{ij} t_{jk} t_{ki}}{U^{2}}\{\cos{\theta_{ijk}} \mathcal{J}_{ik} \mathbf{S}_{k} \cdot(\mathbf{S}_{i}\times \mathcal{J}_{ij} \mathbf{S}_{j})
 \nonumber \\
&+& \sin{\theta_{ijk}} \; \mathbf{n}_{ijk}\cdot[\mathbf{S}_{i}(\mathcal{J}_{ij} \mathbf{S}_{j} \cdot\mathcal{J}_{ik} \mathbf{S}_{k}) 
- \mathcal{J}_{ij} \mathbf{S}_{j} (\mathbf{S}_{i}\cdot \mathcal{J}_{ik} \mathbf{S}_{k})
\nonumber \\
&-&\mathcal{J}_{ik} \mathbf{S}_{k} (\mathbf{S}_{i}\cdot \mathcal{J}_{ij} \mathbf{S}_{j}) + (\mathbf{S}_{i}+\mathcal{J}_{ij} \mathbf{S}_{j}+\mathcal{J}_{ik} \mathbf{S}_{k})/12 ]\}
 \label{eq:Curr0}
 \end{eqnarray}
The current density operator is the same for the other two bonds $jk$ and $ki$, except for the factor ${\bf e}_{ij}$, which has to be replaced by ${\bf e}_{jk}$ and ${\bf e}_{ki}$, respectively.

These results can be easily  extended to the lattice by adding the contributions from each triangular loop that contains 
a given lattice site $j$ in the case of the charge density operator and a given bond $\langle jl \rangle$, in the case of the effective
current-density operator.

\section{Limit of weak spin-orbit coupling  \label{wsocl}}

The SOC is weak when the magnetic moment is provided by a $3d$ transition metal. Therefore, it is relevant to discuss the weak spin orbit coupling limit $\theta_{jl}\ll1$. By keeping terms  up to linear order in $\theta_{jl}$ in Eq.~\eqref{eq:effHami2},  the effective Hamiltonian reduces to
\begin{equation}
{\tilde H} =\sum_{ \langle jl \rangle } J_{jl} \mathbf{S}_{j}\cdot\mathbf{S}_{l}+\mathbf{D}_{ jl }\cdot(\mathbf{S}_{j}\times\mathbf{S}_{l})\label{eq:weakEP}
\end{equation}
where $J_{jl}=4t_{jl}^{2} /U$ and $\mathbf{D}_{jl}=2 J_{jl} \theta_{jl} \mathbf{n}_{jl}$
to  first order in $\theta_{jl}$. For the effective polarization operators of the single triangle shown in Fig.~\ref{fig:SOCsche}, we obtain
\begin{equation}
\mathbf{\tilde P} = \mathbf{\tilde P}_{ij} + \mathbf{\tilde P}_{jk} + \mathbf{\tilde P}_{ki} 
\label{sum3}
\end{equation}
with 
\begin{equation}
\mathbf{\tilde P}_{ij}  = 8 ea \frac{t_{ij} t_{jk} t_{ki}}{U^3}  [ \mathbf{S}_{i}\cdot\mathbf{S}_{j} + {\bf L}_{ij} \cdot \mathbf{S}_{i} \times \mathbf{S}_{j}]  ({\bf e}_{jk}  - {\bf e}_{ki}),
\label{bondP}
\end{equation}
and $\mathbf{L}_{ ij }=2\theta_{ij}\mathbf{n}_{ij} -\theta_{ij}\mathbf{n}_{ij}-\theta_{jk}\mathbf{n}_{jk}-\theta_{ki}\mathbf{n}_{ki}$. $\mathbf{\tilde P}_{jk}$ and $\mathbf{\tilde P}_{ki}$ are obtained from Eq.~\eqref{bondP} by a cyclic permutation of indices $ijk$.
Equations~\eqref{sum3} and \eqref{bondP} coincide with the expressions that were recently derived in Ref.~\onlinecite{arxiv.1113} for studying the effect of SOC on the optical conductivity of Mott insulators.  In the absence of $\mathrm{SU}(2)$ flux ( $\theta_{ijk}\mathbf{n}_{ijk}=0$), the expression reduces to the equation that was originally derived in Ref.~\onlinecite{PhysRevB.78.024402}. 

The new contribution to the electric polarization  is related to the DM interactions on the three bonds of the triangular unit. A finite $\mathrm{SU}(2)$ flux leads to a contribution that is proportional to the vector spin chirality (or spin current) on each bond. This is not surprising on symmetry grounds. The effective charge density operator is a scalar under under the  group of rotations (spin plus orbital degrees of freedom) and it is even under time reversal. Once the lattice anisotropy allows for a vector ${\bf n}_{jl}$ on a given bond $\langle jl \rangle$, we can build a scalar by taking the direct product of  ${\bf n}_{jl}$ with the crossed product between two spins. As we will see below, this new contribution to the electric polarization leads to a new microscopic mechanism for ferroelectricity induced by spiral spin ordering. Another combination that is allowed by symmetry is the product ${\bf n}_{jl} \cdot {\bf S}_{j}$ ${\bf n}_{jl} \cdot {\bf S}_{l}$, which is of order $\theta^2_{jl}$ because it is bilinear in ${\bf n}_{jl}$. Therefore,  although contributions of this form are included  in Eqs.~\eqref{eq:px} and \eqref{eq:py}, they do not appear in the expansions \eqref{sum3} and \eqref{bondP}  up to first order in $\theta_{jl}$.

\begin{figure}
\includegraphics[width=8.5cm]{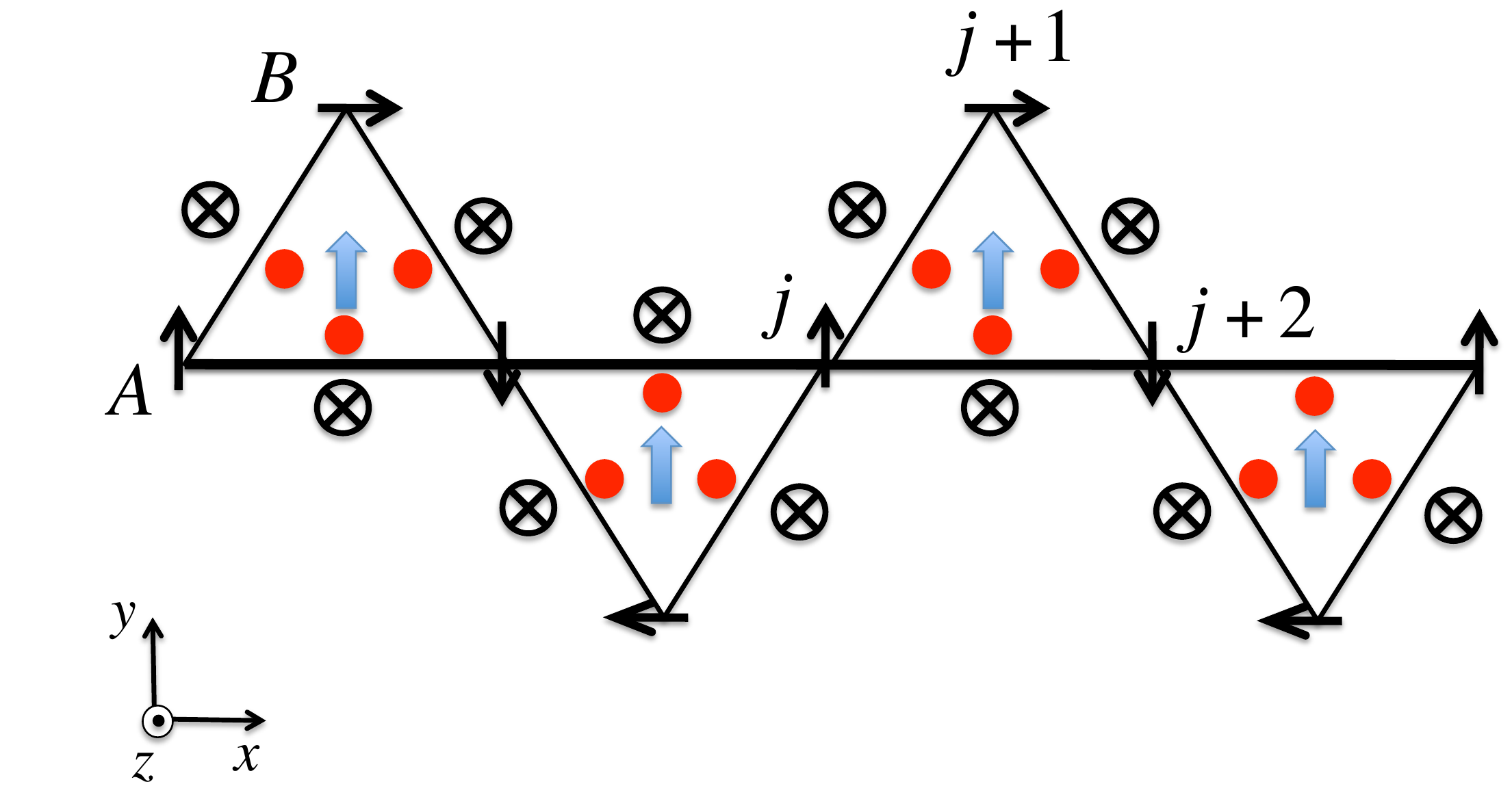}
\caption{\label{fig:sawtooth}  Sawtooth chain with a uniform DM vector parallel to the $z$ axis. The short thin arrows indicate the spin orientation for a cycloidal spiral ordering with wave-vector ${\bf Q} = (\pi/a) {\hat {\bf x}}$. The small circles inside the triangles indicate the positions of ions that mediate the superexchange interactions between magnetic moments. The DM vectors point along the negative $z$ direction (crossed circles) if we circulate anticlockwise around each triangle. The thick arrows indicate the direction of the electric polarization that is induced by the SOC according to Eqs.~\eqref{sum3} and \eqref{bondP}. The exchange interaction is $J$ ($j'$) along the horizontal (oblique) thick (thin) bonds.}
\end{figure}

\subsection{Cycloidal spiral ordering \label{cs}}

To illustrate the potential consequences of the electric polarization component that arises from a finite SOC,  we will first consider the simple  case of a sawtooth chain with uniform DM interactions and DM vectors antiparallel to the $z$ axis.
Figure.~\ref{fig:oxy} illustrates one possible physical realization of this model. Imagine that the hopping between two atoms has a contribution that is mediated by an atom of a different kind. This is a common situation in transition metal oxides, where the magnetic moments are provided by $3d$ orbitals of the transition-metal, but the hopping between transition metals has a contribution that is mediated by the $p$-orbitals of oxygen ions. After integrating out the higher energy orbitals of the intermediary $O^{2-}$ ions, we get a contribution to the hopping term of the Hubbard model, in addition to the one that arises from direct overlap between  the $3d$ orbitals. Based on symmetry arguments,~\cite{Moriya} the  DM vector ${\bf D}_{ij}$ on 
a given bond $\langle ij \rangle$ is proportional to ${\bf x}_{ij} \times {\bf r}_{ij}$: ${\bf D}_{ij} \propto {\bf x}_{ij} \times {\bf r}_{ij}$, where ${\bf x}_{ij}$  is the displacement of the intermediary ion relative to the center of the bond  (see Fig.~\ref{fig:oxy}). It is then clear that the displacements depicted in Fig.~\ref{fig:oxy}  (oxygen ions moving from the bond centers toward the center of the triangles) produce  DM vectors on each bond that are anti-parallel to the $z$-axis: ${\bf D}_{jk} = - (2 J_{jk} \theta_{ijk}/3) {\bf e}_z$ for any bond $\langle j \rightarrow k  \rangle $ oriented  anticlockwise. Correspondingly, the vector potential is  given by $\mathcal{A}_{lk}= t_{lk} e^{-i \theta_{lk} \sigma_{z}}$ for any bond $\langle l \rightarrow k  \rangle $ oriented  counterclockwise.

We  now consider the case of the sawtooth chain that is illustrated in Fig.~\ref{fig:sawtooth}. The horizontal bonds are connected by a hopping amplitude $t_{jk}=t$, whereas  the hopping amplitude on the oblique bonds is $t_{ij}=t_{ki}=t'$. The corresponding exchange interactions are $J_{jk} =J=4 t^2/U$ and $J_{ij}=J_{ki}=J'=4 t'^2/U$ respectively. We will assume that $J > J'/2$ and that the plane of the  saw-tooth chain is an easy-plane. This easy-plane anisotropy can either arise from the symmetric exchange anisotropy terms of  Eq.~\eqref{eq:effHami2}] or  be induced by applying a uniform magnetic field along the  $z$-axis (see Fig.~\ref{fig:sawtooth}).

In the {\it classical limit}, the ground state of the effective spin Hamiltonian is a spiral state:
\begin{eqnarray}
\langle S^{x}_j \rangle  &=& S \sin(Q j a/2 + \phi),
\nonumber \\
\langle S^{y}_j \rangle  &=& S \cos(Q j a/2 + \phi),
\nonumber \\
\langle S^{z}_j \rangle  &=& 0.
\label{stspiral}
\end{eqnarray}
The integer $j$  labels the spins of the saw-tooth chain, $\phi$ is an arbitrary phase, and  $\cos{(Qa/2)}=-J'/2J$. $Q$ is  the amplitude  spiral wave-vector ${\bf Q}= Q {\hat {\bf x}}$.   The spiral ordering described by Eq.~\eqref{stspiral} is not the only ground state in the classical limit of the Hamiltonian under consideration. There is another ground sate in which the spins on the $B$ sites are ferromagnetically aligned. However, this degeneracy can be easily removed by adding an antiferromagnetic exchange interaction between consecutive $B$ sites. We will then assume that the DM interaction does not change this spiral ordering because it is much smaller than the isotropic exchange interactions\footnote{ We note that this assumption does not hold  for the particular case $Q=\pm 4 \pi/3a$ because the spiral ordering is degenerate with another ordering that is favored by the DM interaction. This ordering also consists of a 120 degree structure on each triangle (like the $Q=\pm 4 \pi/3a$ spiral), but with the same vector chirality on the up and down triangles. }..  For simplicity, we will also assume that  $D'/J'=D/J$.

We can now compute the electronic contribution to the electric polarization for the spiral ordering described by Eq.~\eqref{stspiral}. We first observe that 
$\langle {\bf S}_j \cdot {\bf S}_{j+1}  \rangle \neq \langle {\bf S}_j \cdot   {\bf S}_{j+2}  \rangle $. This inequality holds even in absence 
of magnetic ordering because the  $A$ and $B$ sites are not equivalent in the sawtooth chain. 
According to Eqs.~\eqref{sum3} and \eqref{bondP},   this inequality implies that there is a staggered electric polarization parallel or antiparallel to the $y$ axis, simply reflecting the fact that the $A$ and $B$ sites of the saw-tooth chain are not equivalent. In contrast, the SOC contribution to  Eqs.~\eqref{sum3} and \eqref{bondP} leads to a {\it uniform}
contribution to the electric polarization, as indicated by the thick arrows shown in Fig.~\ref{fig:sawtooth}. This uniform polarization arises from the fact that the vector chirality 
is staggered for the spiral ordering. In other words, $\langle {\bf S}_i \times {\bf S}_j \rangle$ has an opposite sign for the up and down triangles if the arrow that connects $i$ and $j$
has the same circulation sense (clockwise or anticlockwise) on both triangles.

Our analysis implies that the cycloidal magnetic ordering, depicted in Fig.~\ref{fig:sawtooth} leads to a net electric polarization in the presence of a finite SOC. The value of the {\it uniform} polarization per triangle is given by
\begin{equation}
\langle \mathbf{\tilde P}_u \rangle= \frac{4\sqrt{3} ea}{J}  S^{2}  \frac{t t'^2}{U^3} [ \sin{(Qa/2)} + \sin{Qa} ] \;
\mathbf{D} \times {\hat{\mathbf Q}}.
\label{eq:spiralEPperT}
\end{equation}
where ${\hat{\mathbf Q}}$ is a unit vector parallel to the wave-vector of the spiral (${\hat{\mathbf Q}}={\hat {\bf x}} $ for the saw-tooth chain of Fig.~\ref{fig:sawtooth}). In the long wavelength limit $Q a \ll 1$, Eq.~\eqref{eq:spiralEPperT} becomes 
\begin{equation}
\langle \mathbf{\tilde P}_u \rangle= \frac{6\sqrt{3} ea^2}{J}  S^{2}  \frac{t t'^2}{U^3} 
\mathbf{D} \times {\mathbf Q}.
\label{eq:spiralEPperTlw}
\end{equation}
This expression is similar to the equation obtained from the so-called "inverse DM" mechanism~\cite{Katsura05,PhysRevB.73.094434,PhysRevLett.96.067601}. However, its origin is completely different. In the "inverse DM" mechanism, the  DM interaction is induced by the spiral ordering via an {\it ionic displacement} that produces the net electric polarization. In contrast, for the "direct DM'' coupling mechanism that we are discussing here, the DM interaction is already present in the paramagnetic phase and it induces a finite {\it electronic polarization} when the system develops cycloidal spiral ordering. Therefore, the new mechanism presented here is the direct (electronic) counterpart of the inverse (ionic) DM mechanism introduced in
Refs.~\onlinecite{Katsura05,PhysRevB.73.094434,PhysRevLett.96.067601}.

The spiral ordering discussed for the sawtooth chain can be naturally extended to an anisotopic   kagome lattice  by stacking sawtooth chains on top of each other (see Fig.~\ref{fig:kagome}). In particular, this implies  that the $J, J'$ Hamiltonian of the anisotropic kagome lattice can be expressed as a sum of partial Hamiltonians over the sawtooth chains that compose this lattice. Because the spiral ordering with wave vector parallel to the thick bond direction minimizes the energy of each partial sawtooth chain Hamiltonian, it also minimizes the energy of the global Hamiltonian for the anisotropic Kagome lattice. Therefore,  the uniform electric polarization given by  Eq.~\eqref{eq:spiralEPperT} is also obtained in this case (see Fig.~\ref{fig:kagome}).

\begin{figure}
\includegraphics[width=8.5cm]{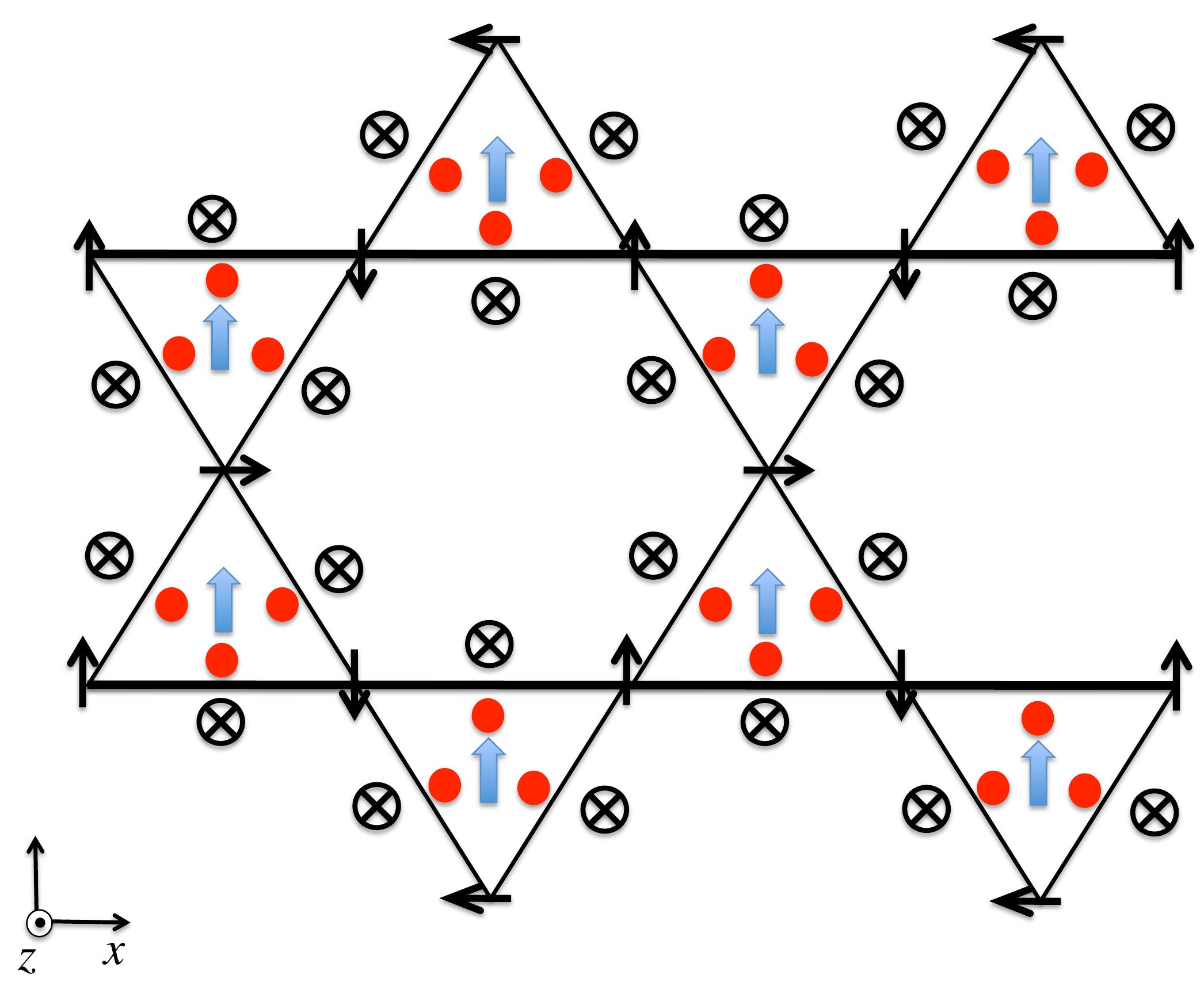}
\caption{\label{fig:kagome}  Anisotropic kagome lattice (the thick horizontal bonds are different from the oblique bonds) with uniform DM vector parallel to the $z$ axis. The short thin arrows indicate the spin orientation for a cycloidal spiral ordering with wave vector ${\bf Q} = (\pi/a) {\hat {\bf x}}$. The other  arrows and symbols have the same meaning as in Fig.~\ref{fig:sawtooth}. }
\end{figure}

It is important to note that quantum fluctuations will in general destroy the $T=0$ spiral ordering of the sawtooth chain, whereas thermal fluctuations will destroy the finite-$T$ spiral ordering of the kagome lattice. This is so because the spiral ordering $\langle {\bf S}_j \rangle \neq 0$ spontaneously breaks the U(1) symmetry of uniform spin rotations along the $z$ axis. However, the vector chirality $\langle {\bf S}_i \times  {\bf S}_j \rangle $, which is the quantity that induces a finite electric polarization, does not break this continuous symmetry (it only breaks the discrete spatial inversion symmetry). Therefore, this quantity will in general survive  together with the electric polarization  in the  presence of finite thermal and/or quantum fluctuations ($\langle {\bf S}_j \rangle = 0$ and $\langle {\bf S}_i \times  {\bf S}_j \rangle \neq 0$). Consequently, Eq.~\eqref{eq:spiralEPperTlw} has to be replaced by
\begin{equation}
\langle \mathbf{\tilde P}_u \rangle= \frac{6\sqrt{3} ea D}{2J}    \frac{t t'^2}{U^3} 
\langle {\bf S}_{\bf r} \times  {\bf S}_{{\bf r} + a \hat {\bf Q}} \rangle   \times {\hat {\mathbf Q}}.
\label{eq:spiralEPperTlwquan}
\end{equation}

\subsection{Proper screw spiral ordering \label{ps}}

We will now consider the alternative situation of proper screw spiral ordering. The proper screw spiral depicted in Fig.~\ref{fig:sawtooth2} can be stabilized by applying a magnetic field ${\bf H}$ parallel to the wave vector of the spiral. Although the uniform component of the magnetic moments is parallel to the applied field (not shown in Fig.~\ref{fig:sawtooth2}), the spiral component is confined to the plane perpendicular to ${\bf H}$. Because we will later consider a two-dimensional case, it is convenient to introduce a spiral wave vector $\mathbf{Q}= Q (\cos\phi_{\mathbf{Q}},\sin\phi_{\mathbf{Q}},0)$ that can point along any direction in the plane ($\phi_{\bf Q}=0$  for the saw-tooth chain of Fig.~\ref{fig:sawtooth2}). The applied magnetic field,
 ${\bf H}= H (\cos\phi_{\mathbf{Q}},\sin\phi_{\mathbf{Q}},0)$, is parallel to $\mathbf{Q}$. 
At the mean field (semi-classical) level, the proper screw spiral spin ordering on the saw-tooth chain depicted in Fig.~\ref{fig:sawtooth2}  is described by the following  equations:
\begin{eqnarray}
\left\langle S^{x}_j \right\rangle &=& S \cos{\alpha}  \sin\phi_{\mathbf{Q}}\sin(Q ja /2+ \phi) + S \sin{\alpha} \cos\phi_{\mathbf{Q}}, 
\nonumber \\
\left\langle S^{y}_j \right\rangle &=& -S \cos{\alpha}  \cos\phi_{\mathbf{Q}}\sin(Q j a/2 +\phi)  + S \sin{\alpha} \sin\phi_{\mathbf{Q}} , 
\nonumber \\
\left\langle S^{z}_j \right\rangle &=& S \cos{\alpha} \cos(Q j a/2 +\phi),
\label{eq:pscspiral}
\end{eqnarray}
which are the counterparts of Eqs.~\eqref{stspiral} for the case of cycloidal spiral ordering. Here $\alpha$ is the canting angle induced by the uniform magnetic field $\mathbf{H}$. 

\begin{figure}
\includegraphics[width=8.5cm]{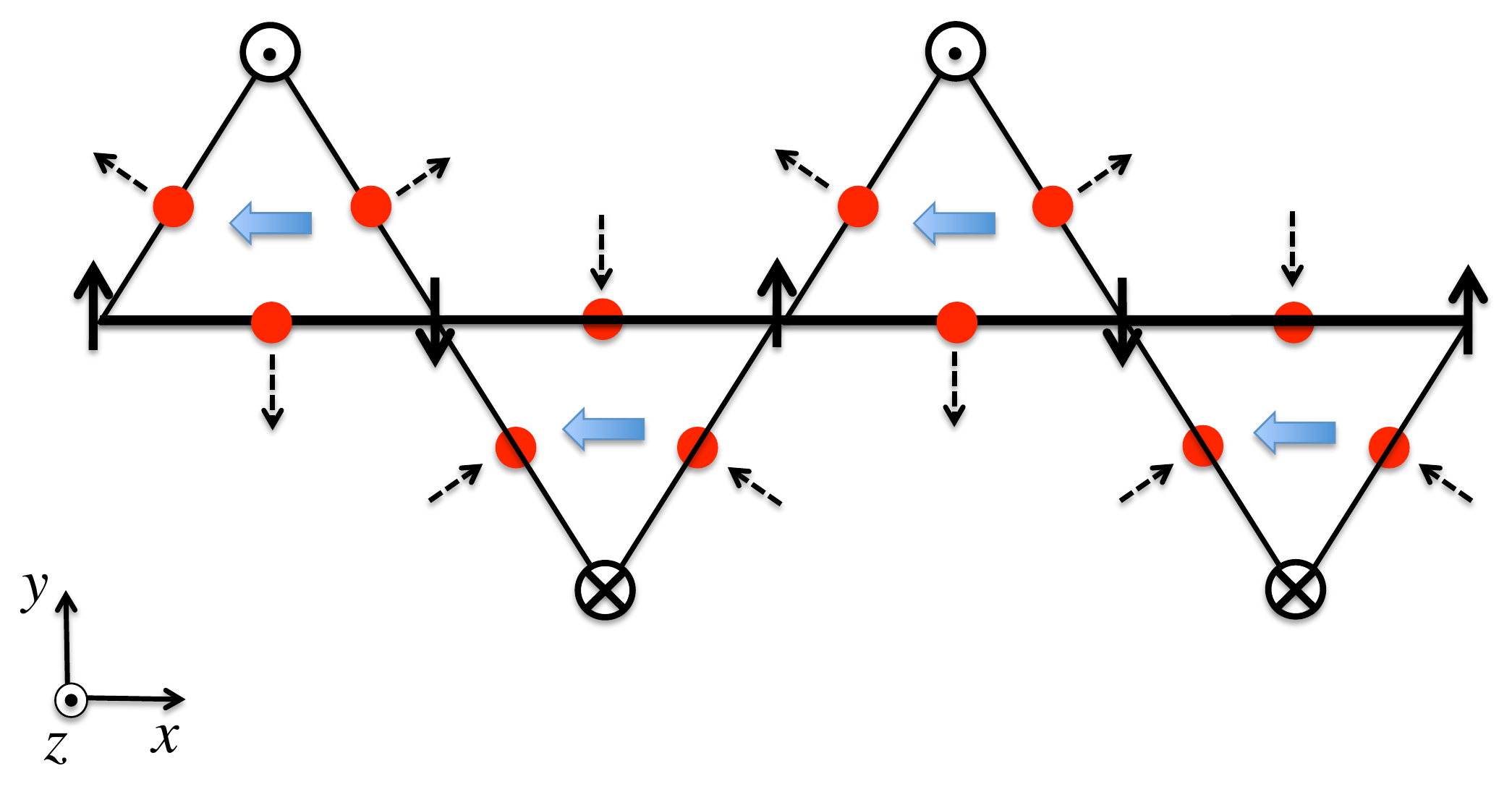}
\caption{\label{fig:sawtooth2}  Sawtooth chain with  DM vectors (dashed arrows) perpendicular to the bond direction.  These DM vectors are induced by intermediate ions (small circles) that are shifted relative to the bond centers along the $z$ direction. The dashed arrows indicate the orientation of the DM vectors when we circulate anticlockwise around each triangle. The short thin arrows indicate the spin orientation for a proper screw spiral ordering with wave vector ${\bf Q} = (\pi/a) {\hat {\bf x}}$.  The thick arrows indicate the direction of the electric polarization that is induced by the SOC according to Eqs.~\eqref{sum3} and \eqref{bondP}.}
\end{figure}

Like in the previous case,  this spiral solution on the sawtooth chain can be extended to an anisotropic kagome lattice  by
stacking the saw-tooth chains on top of each other
(see Fig.~\ref{fig:kagome2}). In this case, the product $Qj$ in Eqs.~\eqref{eq:pscspiral} has to be replaced by ${\bf Q} \cdot {\bf r}_j$, where the vector ${\bf r}_j$ corresponds to the position of  site $j$ in units of $a$. Now we will assume that the DM vectors remain perpendicular to the bonds connecting the two sites, but they are  parallel to the plane of the lattice (see Figs.~\ref{fig:sawtooth2}  and \ref{fig:kagome2}): ${\bf D}_{ij} = D_{ij} {\bf e}_{ij} \times {\hat {\bf z}}$.  This implies that if the DM vectors point outwards in the up triangles, they will point inwards in the down triangles. For instance, in the sawtooth chain and the anisotropic kagome lattice shown Figs.~\ref{fig:sawtooth2}  and \ref{fig:kagome2},  the DM vectors point toward the center of the down triangles and away from the center of the up triangles, when we circulate anticlockwise around each triangle. This implies that the displacement along the $z$ axis of the ions which mediate the superexchange interaction is positive  for up triangles and  negative for down triangles.
In the rest of this section we will see that this in-plane component of the DM interaction produces a uniform component of the electric polarization in the presence of proper screw spiral spin ordering.

The isotropic contribution to the electric polarization given by Eqs.~\eqref{sum3} and \eqref{bondP} does not depend on the polarization plane of the spiral. 
Therefore, like in the cycloidal case, this contribution leads to a staggered electric polarization component  that is already present in the paramagnetic phase. In contrast, it is easy to verify that the SOC to the polarization is the same on every single triangle (up and down). Consequently, to compute the electric polarization per triangle, it is enough to consider a single triangle of the sawtooth chain or the kagome lattice. For simplicity, we will consider an up triangle and denote the three sites with the labels $ijk$, as in Fig.~\ref{fig:SOCsche}.

According to Eq.~\eqref{bondP}, the spin-orbit contribution to ${\tilde {\bf P}}_{ij}$ is
\begin{equation}
\langle  {\tilde {\bf P}}_{ij}^{SO} \rangle = \frac{8  e a S^2 \theta }{\sqrt{3}} \frac{t t'^2}{U^3} \sin{ ({\bf Q} \cdot {\bf r}_{ij})} \;  {\hat {\bf Q}} \cdot {\bf e}_{ij} \times {\hat {\bf z}} \;  {\bf e}_{ij} \times {\hat {\bf z}}, 
\label{sbpsc2}
\end{equation}
where ${\hat {\bf Q}} = (\cos{\phi_{\bf Q}}, \sin{\phi_{\bf Q}},0)$ and ${\bf r}_{ij} = a {\bf e}_{ij}$. Here we have used that  
$\langle\mathbf{S}_{i}\times\mathbf{S}_{j}\rangle=S^{2}\cos^{2}\alpha\sin(\mathbf{Q}\cdot\mathbf{e}_{ij})\hat{\mathbf{Q}}+
S\sin\alpha\hat{\mathbf{Q}}\times(\langle\mathbf{S}_{j}\rangle-\langle\mathbf{S}_{i}\rangle)$ and 
neglected the second term by assuming that the canting angle $\alpha$ is much smaller than one: $\cos{\alpha} \simeq 1$ and $\sin{\alpha} \simeq 0$. We note that
${\bf L}_{ij}= - {\bf D}_{ij}/2J_{ij}$ and that $D_{ij}/J_{ij}= 2 \theta_{ijk}/3= 2\theta/3$ does not depend on the
bond $\langle ij \rangle$, because $D/J =D'/J'$. 
%Note also that ${\bf D}_{ij} = D_{ij} {\bf e}_{ij} \times {\hat {\bf z}}$.

\begin{figure}
\includegraphics[width=8.5cm]{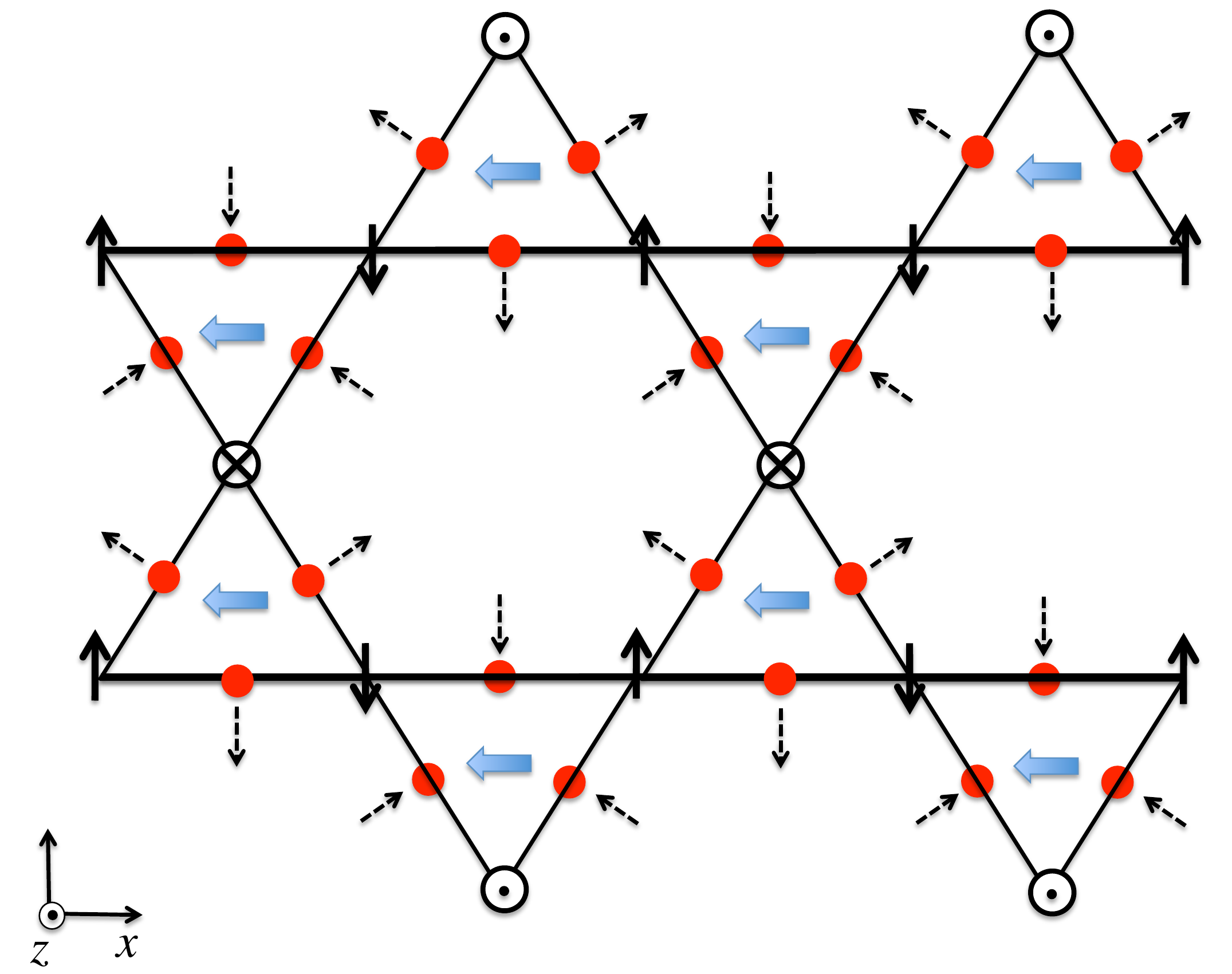}
\caption{\label{fig:kagome2}  Kagome lattice with a DM vector (dashed arrows) perpendicular to the bond direction. The short thin full arrows indicate the spin orientation for a proper screw spiral ordering with wave-vector ${\bf Q} = (\pi/a) {\hat {\bf x}}$. The other arrows have the same meaning as in Fig.~\ref{fig:sawtooth2}. }
\end{figure} 

By using the identities ${\hat {\bf Q}} \cdot {\bf e}_{ij} \times {\hat {\bf z}}  \; {\bf e}_{ij} \times {\hat {\bf z}} = {\hat {\bf Q}} - {\hat {\bf Q}} \cdot {\bf e}_{ij} \; {\bf e}_{ij}$ and
${\bf e}_{ij} = {\hat {\bf Q}} \cdot {\bf e}_{ij} \; {\hat {\bf Q}}  + {\hat {\bf z}} \times {\hat {\bf Q}} \cdot {\bf e}_{ij} \; {\hat {\bf z}} \times {\hat {\bf Q}}$, we obtain
\begin{eqnarray}
\langle  {\tilde {\bf P}}_{ij}^{SO} \rangle &=& \frac{8  e a S^2 \theta }{\sqrt{3}} \frac{t t'^2}{U^3} \sin{ ({\bf Q} \cdot {\bf r}_{ij})} \;
\{ [1- ({\hat {\bf Q}} \cdot {\bf e}_{ij} )^2]{\hat {\bf Q}} 
\nonumber \\
&-& ({\hat {\bf Q}} \cdot {\bf e}_{ij}) \; ({\hat {\bf z}} \times {\hat {\bf Q}}\cdot {\bf e}_{ij}) 
\; {\hat {\bf z}} \times {\hat {\bf Q}} \}, 
\label{sbpsc3}
\end{eqnarray}
In this way, we have divided $\langle  {\tilde {\bf P}}_{ij}^{SO} \rangle$ into a component $\langle  {\tilde {\bf P}}_{ij, \parallel}^{SO} \rangle$,
which is parallel to the wave-vector ${\bf Q}$, and a component, $\langle  {\tilde {\bf P}}_{ij, \perp}^{SO} \rangle$, which is perpendicular to 
${\bf Q}$ (parallel to ${\hat {\bf z}} \times {\hat {\bf Q}}$)
\begin{eqnarray}
\langle  {\tilde {\bf P}}_{ij, \parallel}^{SO} \rangle &=& \frac{8  e a S^2 \theta }{\sqrt{3}} \frac{t t'^2}{U^3} \sin{ ({\bf Q} \cdot {\bf r}_{ij})} \;
 [1- ({\hat {\bf Q}} \cdot {\bf e}_{ij} )^2]
 \nonumber \\
\langle  {\tilde {\bf P}}_{ij, \perp}^{SO} \rangle &=&  \frac{8  e a S^2 \theta }{\sqrt{3}} \frac{t t'^2}{U^3} \sin{ ({\bf Q} \cdot {\bf r}_{ij})} 
 \; {\hat {\bf Q}} \cdot {\bf e}_{ij} \;  {\hat {\bf Q}} \times {\hat {\bf z}}  \cdot {\bf e}_{ij} 
 \label{parperp}
\end{eqnarray}
The  uniform polarization per triangle $\langle  {\tilde {\bf P}}_{u} \rangle$ is obtained by adding the contributions from the three bonds of the triangle. The value of $\langle  {\tilde {\bf P}}_{u} \rangle$ for arbitrary $Q$ is given by,
\begin{eqnarray}
\langle  {\tilde {\bf P}}_{u, \parallel}\rangle &=& \frac{8  e a S^2 \theta }{\sqrt{3}} \frac{t t'^2}{U^3} 
\{ \sin{ ({\bf Q} \cdot {\bf r}_{ij})} \;  [1- ({\hat {\bf Q}} \cdot {\bf e}_{ij} )^2] 
\nonumber \\
&+& \sin{ ({\bf Q} \cdot {\bf r}_{jk})} \;  [1- ({\hat {\bf Q}} \cdot {\bf e}_{jk} )^2]  
\nonumber \\
&+& \sin{ ({\bf Q} \cdot {\bf r}_{ki})} \;  [1- ({\hat {\bf Q}} \cdot {\bf e}_{ki} )^2]   \}
 \nonumber \\
\langle  {\tilde {\bf P}}_{u, \perp} \rangle &=&  \frac{8  e a^2 S^2 \theta }{\sqrt{3}} \frac{t t'^2}{U^3} [\sin{ ({\bf Q} \cdot {\bf r}_{ij})} 
 \; {\hat {\bf Q}} \cdot {\bf e}_{ij} \;  {\hat {\bf Q}} \times {\hat {\bf z}}  \cdot {\bf e}_{ij} 
 \nonumber \\
 &+& 
 \sin{ ({\bf Q} \cdot {\bf r}_{jk})} 
 \; {\hat {\bf Q}} \cdot {\bf e}_{jk} \;  {\hat {\bf Q}} \times {\hat {\bf z}}  \cdot {\bf e}_{jk} 
 \\ \nonumber 
 &+& \sin{ ({\bf Q} \cdot {\bf r}_{ki})} 
 \; {\hat {\bf Q}} \cdot {\bf e}_{ki} \;  {\hat {\bf Q}} \times {\hat {\bf z}}  \cdot {\bf e}_{ki} ]
 \label{parperp2}
\end{eqnarray}
After taking the long wavelength limit $ Q a \ll 1 $ and making the  approximation $\sin{ ({\bf Q} \cdot {\bf r}_{ij})}  \simeq {\bf Q} \cdot {\bf r}_{ij} $, we obtain
\begin{eqnarray}
\langle  {\tilde {\bf P}}_{u} \rangle \simeq - \frac{6  e a^2 S^2 \theta Q}{\sqrt{3}} \frac{t t'^2}{U^3} 
[ \cos{3\phi_{\bf Q}} \;  {\hat {\bf Q}}  - \sin{3\phi_{\bf Q}}  \; {\hat {\bf z}} \times {\hat {\bf Q}} ]
 \label{parperp3}
\end{eqnarray}
This result implies that the uniform electric polarization induced by the SOC is  antiparallel to ${\bf Q}$ (for $D_{ij}>0$) when ${\bf Q}$ is parallel to a bond direction (see Fig.~\ref{fig:kagome2}).  In addition, $\langle  {\tilde {\bf P}}_{u} \rangle$  rotates clockwise by an angle $3\phi_{\bf Q}$  in the reference frame attached to ${\bf Q}$, when ${\bf Q}$ rotates anti-clockwise by an angle $\phi_{\bf Q}$. This implies that $\langle  {\tilde {\bf P}}_{u} \rangle$ is orthogonal to ${\bf Q}$ when ${\phi}_{\bf Q}$ is an odd multiple of 
$\pi/6$. It is important to clarify that the wave vector ${\bf Q}$ is always parallel to the thick bond direction (see Fig.~\ref{fig:kagome2}) for the case of the {\it anisotropic} Kagome lattice. However, proper-screw spiral orderings  with  different ${\bf Q}$ directions could in principle appear in {\it isotropic} Kagome lattices with longer range exchange interactions. This is the reason why we have contemplated the more general case in the derivation of Eq.~\eqref{parperp3}. 

The  angular dependence of the electric polarization given by  Eq.~\eqref{parperp3} is the same as the one  predicted by Arima's mechanism~\cite{JPSJ.76.073702} and measured in the 
the triangular-lattice helimagnet $\mathrm{MnI}_2$\cite{PhysRevLett.106.167206}. Although the microscopic mechanism that led to 
Eq.~\eqref{parperp3} is completely different, the angular dependence of the electric polarization induced by a proper screw spiral is determined by symmetry. Consequently, it is independent of the particular microscopic mechanism that couples the magnetic ordering 
to the uniform electric polarization.

It is important to note that there is another qualitative difference between the electronic mechanism that we have  derived in secs
\ref{cs} and \ref{ps} for cycloidal and proper screw spirals, and the seemingly related inverse DM and Arima's mechanisms, which are based on ionic displacements. The electronic mechanism  arise from loop (of three sites) contributions, whereas the ionic mechanisms arises from bond (two sites) contributions. As is clear from the last factor, $({\bf e}_{jk}  - {\bf e}_{ki})$, of Eq.~\eqref{bondP}, if there is spiral ordering on the triangle $ijk$, the sign of the electric polarization depends on whether the site $i$ is above or below the bond $jk$. Consequently, single-${\bf Q}$ spiral orderings on  triangular lattices do not produce  a net electronic polarization of the type discussed in this paper because the contributions from up and down triangles cancel with each other (see Fig.~\ref{fig:trianlatt}). In contrast, as is discussed in Ref.~\onlinecite{JPSJ.76.073702}, single-${\bf Q}$ 
spiral orderings on triangular lattices produce net ionic contributions to the electric polarization via the inverse DM and Arima's mechanisms. There is no cancellation in this case because the contribution from each bond $\langle ij \rangle$ does not depend on the  position of a third site $k$. 
The situation is different for the sawtooth  and  kagome lattices  because  each bond belongs to a single triangle in these lattices. We emphasize that a net (uniform) component of the  electric polarization arises from a staggered component of the vector chirality, i.e., opposite vector chirality on the up and down triangles. 

\begin{figure}
\includegraphics[width=8.5cm]{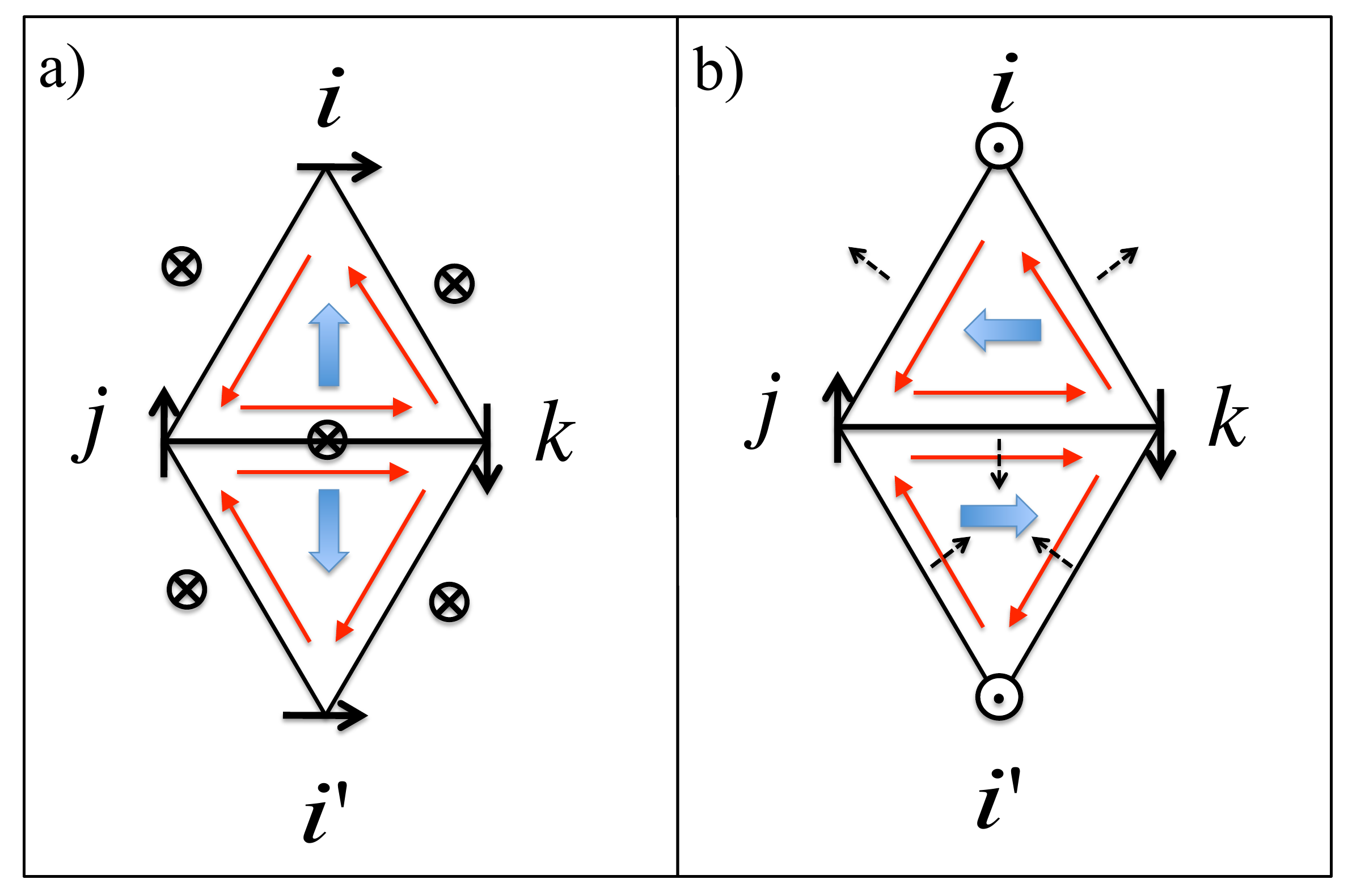}
\caption{\label{fig:trianlatt}  There is no net electronic polarization for spiral magnetic ordering on a triangular lattice because 
the contributions from the up and down triangles have opposite signs. The figure shows the electric polarizations induced by the finite 
SOC on a pair of up and  down triangles for (a) cycloidal spiral ordering and (b)  proper screw spiral ordering. The arrows that are parallel to
the bonds indicate the bond orientation for the directions of the DM vectors that are shown in the figure. The rest of the arrows have the same meaning as in the previous figures.}
\end{figure}

\section{Conclusions \label{conc}}

To summarize, we have studied the charge effects induced by SOC deep inside the Mott insulating regime. The SOC generates new contributions to the effective charge density operator, which are proportional to the vector spin chirality $\langle {\bf S}_i \times {\bf S}_j \rangle$ on bonds that belong to the same triangle as the site under consideration. This simple observation implies that SOC can potentially induce a net electric polarization for magnetic orderings that exhibit a spontaneous vector spin chirality $\langle {\bf S}_i \times {\bf S}_j \rangle \neq 0$. Although this chiral ordering can appear in low-dimensional systems, even in absence of magnetic ordering ($\langle {\bf S}_i \rangle=0$),~\cite{Ivar08,Khaled09,Kato10} the most common realization for higher dimensional systems corresponds to spiral magnetic ordering: $\langle {\bf S}_i \times {\bf S}_j \rangle  \simeq \langle {\bf S}_i \rangle \times \langle {\bf S}_j \rangle \neq 0$ . 
Here we have shown that SOC leads to a net electronic polarization for cycloidal and proper-screw spiral orderings on the  sawtooth and 
anisotropic Kagome lattices. An important property of these lattices is that each bond belongs to only one triangle. This property avoids the cancellation between contributions from opposite triangles  that occurs in triangular lattices (see Fig.~\ref{fig:trianlatt}).
Like for the isotropic case, frustration is an essential ingredient for having a net electronic polarization. This is so because the effective charge density operator is odd under a particle-hole transformation that changes the sign of the hopping amplitudes, i.e., it 
only has contributions from  loops  which are closed by an odd number of hopping amplitudes.~\cite{PhysRevB.78.024402,Lin13}

The SOC-induced magnetoelectric coupling described in this paper has some resemblances with known ionic-based mechanisms that have been proposed for spiral orderings in the previous literature.~\cite{Katsura05,PhysRevB.73.094434,PhysRevLett.96.067601,JPSJ.76.073702} This is not surprising given that the relation between the electric polarization and the magnetic ordering is strongly constrained by symmetry. Although the dimensionless parameter $\theta$, which measures the strength of the SOC, appears both in the ionic and in the electronic contributions to the electric polarization field induced by spiral ordering, the other small dimensionless parameter is different in both cases.  The ionic contribution is  proportional to the ratio between the exchange interaction and the energy of the optical mode associated with the ionic displacement. In contrast, the electronic contribution is proportional to $8 t^3/U^3$. Therefore,
the ionic contribution should dominate in the strong-coupling limit $t/U \to 0$, whereas the electronic contribution  becomes dominant when moving towards the intermediate coupling regime.  

Finally, it is important to mention that other contributions to the electronic polarization  also arise when moving away from the weak SOC limit. This regime is relevant for heavy magnetic ions, such as $4d$ and $5d$ transition metals, or lanthanide ($4f$) and actinide ($5f$) elements. We note that these ions  also contain several valence orbitals, implying that they have to be described with multi-band Hubbard models, instead of the simplified single-band model that we have considered here. The spirit of the present paper is simply to demonstrate that SOC leads to new electronic contributions to the electric polarization field induced by certain magnetic orderings. The results that we have derived for a minimal single-band Hubbard model can be naturally extended to the more complex models that describe the low-energy physics of each particular magnet.

\begin{acknowledgements}

We would like to thank  Y.-H. Liu, H.-B. Chen, S.-Z. Lin and Y. Kamiya for  helpful discussions. S. Zhu and Y.-Q. Li are supported by NSFCs (Grants No.11074216 and No. 11274272) and the Fundamental Research Funds for the Central Universities of China. Work at LANL was performed under the auspices of the U.S. DOE Contract No. DE-AC52-06NA25396 through the LDRD program. The work of C.D.B. was supported, in part, by the National Science Foundation under
Grant No. PHYS-1066293 and the hospitality of the Aspen Center for Physics.

\end{acknowledgements}

\begin{appendix}

\section{Effective charge density and current in the presence of SOC}

By applying the canonical transformation \eqref{canon} to the density and current density operators \eqref{dens} and \eqref{cdens}, we obtain expressions for the effective  operators in the low-energy sector of the Mott insulator.
The following identity holds for the effective density operator:
\[
\tilde{n}_{i}=1+P_{s}\{T_{-1}T_{0}\left[T_{1},n_{1}\right]+\left[T_{1},n_{i}\right]T_{0}T_{1}\}P_{s}/U^3.
\]
The effective current density operator is given by
\[
\tilde{\bf I}_{ij}=P_{s}\{T_{-1}T_{0} {\bf I}_{ij}+T_{-1}{\bf I}_{ij}T_{1}+ {\bf I}_{ij}T_{0}T_{1}\}/U^2.
\]
For the single triangle  shown in Fig.~\ref{fig:SOCsche}, we obtain
\begin{equation}
\delta {\tilde n}_{i}=\delta {\tilde n}_{i,jk}+\delta {\tilde n}_{i,kj}\label{eq:charff}
\end{equation}
with $\delta n_{i,jk}=t_{ij}t_{jk}t_{ki}/U^3[  \mathcal{P}_{ikj}-\mathcal{P}_{jki}+\mathcal{Q}_{kji}-\mathcal{Q}_{kij}+\mathrm{H.c.}]$. Here $\mathcal{P}_{i_1i_2i_3}=\chi_{i_1i_2}n_{i_2\uparrow}n_{i_2\downarrow}\chi_{i_2i_3}n_{i_3\uparrow}n_{i_3\downarrow}\chi_{i_3i_1}$ and $\mathcal{Q}_{i_1i_2i_3}=\chi_{i_1i_2}(1-n_{i_1\uparrow}n_{i_1\downarrow})\chi_{i_3i_1}n_{i_2\uparrow}n_{i_2\downarrow}\chi_{i_2i_3}$ where $\chi_{i_1i_2}= {\bf c}_{i_1}^{\dagger} \mathcal{A}_{i_1i_2} {\bf c}_{i_2}$. 
The effective current density operator on the bond $\langle ij \rangle$ can be expressed as
\begin{eqnarray}
\tilde{{\bf I}}_{ij} & =&\frac{i {\bf e}_{ij}}{\hbar}t_{ij}t_{jk}t_{ki}/U^2[\mathcal{P}_{ikj}+\mathcal{P}_{jik}+\mathcal{P}_{kji}
 \nonumber \\
&+&\mathcal{Q}_{kji}+\mathcal{Q}_{ikj}+\mathcal{Q}_{jik} -\mathrm{H.c.}].\label{eq:currentff}
\end{eqnarray}
%with
%\begin{equation}
%\chi_{ik} \chi_{kj} \chi_{ji} = {\bf c}_{i}^{\dagger} \mathcal{A}_{ik} {\bf c}_{k}  {\bf c}_{k}^{\dagger} \mathcal{A}_{ki} \mathcal{A}_{ik} \mathcal{A}_{kj} \mathcal{A}_{ji} \mathcal{A}_{ij} {\bf c}_{j} {\bf c}_{j}^{\dagger} \mathcal{A}_{ji} {\bf c}_{i}.
%\end{equation}
By using the definition $\mathcal{A}_{ik}\mathcal{A}_{kj}\mathcal{A}_{ji} \equiv e^{i\theta_{ijk}\mathbf{n}_{ijk}\cdot\mathbf{\sigma}}$,  we can expand the above equation as 
\begin{equation}
\delta {\tilde n}_{i}=\cos\theta_{ijk}\delta {\tilde n}_{i}^0+i\sin\theta_{ijk}n_{ijk}^\mu\delta {\tilde n}_{i}^\mu,
\label{eachn}
\end{equation}
and
\begin{equation}
\tilde{{\bf I}}_{ij}=\cos\theta_{ijk}\tilde{{\bf I}}_{ij}^0+i\sin\theta_{ijk}n_{ijk}^\mu\tilde{{\bf I}}_{ij}^\mu,
\label{eachcurr}
\end{equation}
where we are using Einstein's convention of summation over repeated index $\mu$=$(x,\,y,\,z)$.
%\begin{eqnarray}
%\chi_{ik} \chi_{kj} \chi_{ji} &=& \cos\theta_{ijk}{\bf c}_{i}^{\dagger} \mathcal{A}_{ik} {\bf c}_{k}  {\bf c}_{k}^{\dagger} \mathcal{A}_{ki} \mathcal{A}_{ij} {\bf c}_{j} {\bf c}_{j}^{\dagger} \mathcal{A}_{ji} {\bf c}_{i} 
%\nonumber \\
%&+& i\sin\theta_{ijk}{\bf c}_{i}^{\dagger} \mathcal{A}_{ik} {\bf c}_{k}  {\bf c}_{k}^{\dagger} \mathcal{A}_{ki} (\mathbf{n}_{ijk}\cdot\mathbf{\sigma})\mathcal{A}_{ij} {\bf c}_{j} {\bf c}_{j}^{\dagger} \mathcal{A}_{ji} {\bf c}_{i}.
%\nonumber \\
%\label{eq:caldetail}
%\end{eqnarray}
Eqs.~\eqref{eq:chargedis}-~\eqref{eq:Curr0} are obtained by calculating each component in Eqs.~\eqref{eachn} and~\eqref{eachcurr}. We can easily obtain $\delta {\tilde n}_{i}^0$ and $\tilde{{\bf I}}_{ij}^0$ by generalizing the result in Ref.~\onlinecite{PhysRevB.78.024402} via  ${\bf S}_i\rightarrow{\bf S}_i$, ${\bf S}_j\rightarrow\mathcal{J}_{ij}{\bf S}_j$ and ${\bf S}_k\rightarrow\mathcal{J}_{ik}{\bf S}_k$ (note that these two contributions are the only ones that survive when the product $\mathcal{A}_{ik} \mathcal{A}_{kj} \mathcal{A}_{ji}$ is proportional to the identity matrix ). 
Because $\delta{\tilde n}_{i}$ and ${\tilde {\bf I}}_{ij}$ are scalars under global spin rotations  (spin and orbital), we just need to calculate only one component (the other two are determined by invoking rotational invariance). 
We consider the case  $\theta_{ijk}=\frac{\pi}{2}$ and ${\bf n}_{ijk}={\bf z}$ by assuming that $\mathcal{A}_{kj}=i\sigma_z$ and both $\mathcal{A}_{ij}$ and $\mathcal{A}_{ik}$ are proportional to the identity matrix. From Eqs.~\eqref{eq:charff} and~\eqref{eq:currentff}, we obtain
\begin{eqnarray}
\delta {\tilde n}_{i}^z&=&-i8\frac{t_{ij}t_{jk}t_{ki}}{U^3}[{\bf S}_i\times {\bf S}_j +{\bf S}_k\times {\bf S}_i-2{\bf S}_j\times {\bf S}_k]^z,
\nonumber\\
\tilde{{\bf I}}_{ij}^z&=&i24\frac{t_{ij}t_{jk}t_{ki}}{U^2}[{\bf S}_i({\bf S}_j\cdot{\bf S}_k)-{\bf S}_j({\bf S}_k\cdot{\bf S}_i)-{\bf S}_k({\bf S}_i\cdot{\bf S}_j)
\nonumber\\
&+& ({\bf S}_i+{\bf S}_j+{\bf S}_k)/12]^z.
\end{eqnarray}
For a general $\mathcal{A}_{ij}$ and $\mathcal{A}_{ik}$, we need to replace  ${\bf S}_j$ and ${\bf S}_k$  by 
$\mathcal{J}_{ij}{\bf S}_j$ and $\mathcal{J}_{ik}{\bf S}_k$, respectively, in the above expressions for $\delta {\tilde n}_{i}^z$ and $\tilde{{\bf I}}_{ij}^z$. In this way we arrive to Eqs.~\eqref{eq:chargedis}-~\eqref{eq:Curr0}.

\end{appendix}

\bibliographystyle{apsrev}
\bibliography{multifer_Tri}

\end{document}